\documentclass{ws-ijmpd2}
\def\ti#1{}

\def\al{\alpha}
\def\be{\beta}
\def\ga{\gamma}
\def\de{\delta}
\def\ep{\epsilon}

\def\ze{\zeta}
\def\et{\eta}
\def\th{\theta}

\def\ka{\kappa}
\def\la{\lambda}

\def\si{\sigma}

\def\ta{\tau}

\def\ps{\psi}
\def\om{\omega}
\def\Ga{\Gamma}

\def\Om{\Omega}
\def\mn{{\mu\nu}}

\def\cl{{\cal L}}
\def\fr#1#2{{{#1} \over {#2}}}

\def\ol#1{\overline{#1}}
\def\lrvec#1{ \stackrel{\leftrightarrow}{#1} }
\def\prt{\partial}
\def\ket#1{|{#1}\rangle}
\def\bra#1{\langle{#1}|}
\def\hmf{ \widehat{m}_F }
\def\tmf{ \widetilde{m}_F }

\def\pr#1{{#1}^\prime}
\def\tb{\tilde{b}}
\def\td{\tilde{d}}
\def\tg{\tilde{g}}
\def\tc{\tilde{c}}

\def\a{$a_\mu$}
\def\b{$b_\mu$}
\def\c{$c_{\mu\nu}$}
\def\d{$d_{\mu\nu}$}
\def\e{$e_\mu$}
\def\f{$f_\mu$}
\def\g{$g_{\la\mu\nu}$}
\def\H{$H_{\mu\nu}$}

\def\half{{\textstyle{1\over 2}}}
\def\frac#1#2{{\textstyle{{#1}\over {#2}}}}
\def\lsim{\mathrel{\rlap{\lower4pt\hbox{\hskip1pt$\sim$}}
    \raise1pt\hbox{$<$}}}
\def\gsim{\mathrel{\rlap{\lower4pt\hbox{\hskip1pt$\sim$}}
    \raise1pt\hbox{$>$}}}
\def\sqr#1#2{{\vcenter{\vbox{\hrule height.#2pt
         \hbox{\vrule width.#2pt height#1pt \kern#1pt
         \vrule width.#2pt}
         \hrule height.#2pt}}}}

\def\etal {{\it et al.}}

\newcommand{\beq}{\begin{equation}}
\newcommand{\eeq}{\end{equation}}
\newcommand{\bea}{\begin{eqnarray}}
\newcommand{\eea}{\end{eqnarray}}
\newcommand{\rf}[1]{(\ref{#1})}

\def\codt{\cos{\om_s T_s}}
\def\sodt{\sin{\om_s T_s}}
\def\ctodt{\cos{2\om_s T_s}}
\def\stodt{\sin{2\om_s T_s}}

\def\kaf{k_{AF}}
\def\kf{k_{F}}
\def\kfi{(k_{F})_{\ka\la\mu\nu}}

\begin{document}

\title{PROBING RELATIVITY USING SPACE-BASED EXPERIMENTS
\footnote{Proceedings of invited talk at International Workshop,
``From Quantum to Cosmos: Fundamental Physics Research in Space,''
Warrenton, VA, USA, May 22-24, 2006.}}

\author{NEIL RUSSELL
}

\address{Physics Department,
Northern Michigan University,
\\
Marquette, MI 49855, USA}

\maketitle

\begin{abstract}
An overview of space tests searching for small deviations from
special relativity arising at the Planck scale is given.
Potential high-sensitivity space-based experiments include ones with
atomic clocks, masers, and electromagnetic cavities. We show that a
significant portion of the coefficient space in the Standard-Model
Extension, a framework that covers the full spectrum of possible
effects, can be accessed using space tests.
Some remarks on Lorentz violation in the gravitational sector are
also given.
\end{abstract}


\section{Introduction}
The Standard Model coupled to General Relativity
is considered to be the best existing physical theory of nature.
It is thought to be the
effective low-energy limit of an underlying fundamental theory
that unifies the gravity and matter sectors at the Planck scale.
This underlying theory may well include
Lorentz violation, detectable in experiments
with appropriate types and levels of sensitivity.
These nonstandard effects can be described for practical purposes
using effective field theory.\cite{Kostelecky:1994rn}
If one takes the General-Relativity coupled Standard Model
and adds appropriate terms that involve operators for Lorentz violation,
the result is the Standard-Model Extension (SME),
which has provided a framework for Lorentz testing
for more than a decade.
Fundamental theories describing such violation could
involve string theory\cite{strings}
and spontaneous symmetry breaking.\cite{slv1,slv2}

The Minkowski-spacetime limit of the SME\cite{SME:minkowski}
has been examined in several dozen experiments.
Theoretical aspects of the photon physics of the SME have been
looked at,\cite{km1,km2,em:theor}
one of these being
radiative corrections.\cite{radiative_corr}
Experimental investigations of electromagnetism of the SME
have considered microwave and optical cavities,\cite{photon:cavities}
Doppler-shift experiments,\cite{Cotter:2006hk}
and Cerenkov radiation.\cite{cerenkov}
Work aimed at investigating the electron physics in the SME includes
studies and measurements of electron coefficients
made using Penning traps,\cite{Penning:theory,Penning:expt}
torsion pendula,\cite{torsion:theor,torsion:exp}
and scattering physics\cite{electron:scattering}.
Lorentz-violating effects involving protons and neutrons
have been investigated,\cite{cc,protneut}
and a variety of tests involving clock-comparison
experiments have been performed.\cite{cc:expt}
Similar tests with antihydrogen\cite{hbar}
may be possible in the near future.
This proceedings will focus on the
possibility of testing Lorentz symmetry
in the space environment.\cite{cc:space}
The SME physics associated with muons
has been studied
and experiments performed.\cite{muons}
Similarly, SME physics has been researched for
neutrinos,\cite{neutrino}
the Higgs,\cite{higgs}
and baryogenesis and nucleosynthesis.\cite{baryogen}
Several studies have been done with neutral mesons.\cite{mesons}
The relationship of the SME to noncommutative geometry
has been uncovered,\cite{noncomm}
and numerous other aspects have been investigated.\cite{sme:other}
Several general reviews of the SME are available.\cite{SMEreviews}

In this proceedings article of the
international workshop ``From Quantum to Cosmos: Fundamental Physics Research in Space''
(held May 22-24, 2006 in Warrenton, Virginia, USA),
we first consider the use of clock-comparison experiments
for measuring the coefficients for Lorentz violation
in the Minkowski limit of the SME.
The clocks referred to here take several forms,
including atomic clocks, masers, and electromagnetic cavity oscillators.
We also discuss the pure-gravity sector of the SME.\cite{gr1,gr2,gr3,gr4}

\section{Fermions and the Minkowski limit of the SME}

The lagrangian describing a spin-$\half$ Dirac fermion $\ps$ of mass $m$
in the presence of Lorentz violation is:
\cite{SME:minkowski}
\beq
\cl = \frac{1}{2}i \ol{\ps} \Ga_\nu \lrvec{\prt^\nu} \ps
   - \ol{\ps} M \ps
\quad ,
\label{lagr}
\eeq
where
\bea
M &:=& m + a_\mu \ga^\mu + b_\mu \ga_5 \ga^\mu
   + \half H_\mn \si^\mn
\quad \mbox{and}
\label{M} \\
\Ga_\nu &:=& \ga_\nu + c_\mn \ga^\mu + d_\mn \ga_5 \ga^\mu
+ e_\nu + i f_\nu \ga_5 + \half g_{\la \mu \nu} \si^{\la \mu}
\quad .
\label{Gam}
\eea
In this expression,
the conventional Lorentz-symmetric case is contained in
the first terms in the expressions for $M$ and $\Ga_{\nu}$ above.
The other terms contain conventional Dirac matrices
$\left\{1, \ga_5, \ga^\mu, \ga_5\ga^\mu, \si^{\mu\nu} \right\}$
and coefficients for Lorentz violation
\a, \b, \c, \d, \e, \f, \g, and \H.
Various possible mechanisms can be envisaged for an underlying theory
giving rise to such coefficients,
including spontaneous symmetry breaking.\cite{slv1,slv2}
The coefficients \c\ and \d\ are traceless,
while \H\ is antisymmetric
and \g\ is antisymmetric in its first two indices.
The terms in the equation for $M$ have dimensions of mass,
and those in the equation for $\Ga_{\nu}$ are dimensionless.

The coefficients for Lorentz violation appearing in the lagrangian
can be thought of as fixed geometrical background objects
in spacetime.
Thus,
any experiment rotating in space could detect
time-dependent projections of these geometric quantities.
Boost-dependent effects could in principle be
detected by considering two identical experiments with differing
velocity vectors.
The general approach to finding Lorentz violation
is therefore to compare identical experiments
with differing rotations and boosts.
Equivalently,
one can seek time dependence in a single experiment
as it rotates in space.
The SME shows that Lorentz symmetry is violated
under these `particle transformations'
where the entire experimental configuration
moves relative to another one, or to itself.
Perturbation theory can be used to calculate
the effects of the coefficients for Lorentz violation,
since they are known to be small.
In the case of atomic clocks,
various simplifying assumptions are made,
and it is then possible to calculate the
energy-level corrections that affect the frequency of the clock.

In contrast,
`observer transformations'
preserve Lorentz symmetry.
This is important since it guarantees
agreement between different observers:
different experimenters observing one experimental system from
differently boosted or rotated inertial reference frames will
find that the components of the parameters
\a, \b, \c, \d, \e, \f, \g, and \H\
transform like conventional tensors under
Lorentz transformations.

\section{Clock-comparison Experiments}
The basic function of an atomic clock is to produce
a stable frequency based on an atomic energy-level transition.
In common configurations,
there is a quantization axis defined by a magnetic field.
If the third coordinate axis of the laboratory reference frame is defined
to run parallel to this axis,
then the output frequency is a function of this magnetic-field component:
$f(B_3)$.

One of the key issues relating to stability is the reduction of
this dependence on the magnetic field, which will drift over time.
In the general SME formalism,
the output frequency is of the form
\beq
\om = f(B_3) + \de \om \ ,
\label{om}
\eeq
where the small correction $\de\om$ carries all the Lorentz-violating contributions
to the output frequency.
It can contain terms that are orientation dependent,
an example being the dot product of the spatial part of
$b^{\mu}$ and $\vec{B}$.
In addition, $\de\om$ can depend on the boost velocity of the clock.
These effects arise because of the motion of the laboratory reference frame.
Therefore,
searching for Lorentz violation requires a detailed knowledge of the motion of the
laboratory relative to a reference frame that is known to be inertial
over the life of the experiment.

A number of simplifying assumptions
must be made to perform calculations
of the effects of Lorentz violation on atomic
clocks, due to the complexity of these
atomic systems.
The hamiltonian is split into a conventional part
describing the atom within the chosen nuclear model,
and a perturbative part containing the coefficients for Lorentz violation.
The perturbative hamiltonian has separate terms for
each proton, electron, and neutron, indexed by the letter $w$:
\beq
\pr{h}=\sum_w\sum_{N=1}^{N_w} \de {h}_{w,N}
\quad .
\label{hprime}
\eeq
Here, the atom or ion $w$ has $N_w$ particles
of type $w$,
and $\de {h}_{w,N}$ is the Lorentz-violating correction
for the $N$th particle of type $w$.
Each of the three particle species in the atom has
a set of Lorentz-violation parameters,
so a superscript $w$ must be placed on each of the parameters
\a, \b, \c, \d, \e, \f, \g, and \H.

The perturbations in the energy levels due to Lorentz violation
are calculated by finding the expectation value of the
hamiltonian $\pr{h}$.
Usually,
the total angular momentum $\vec F$ of the atom or ion,
and its projection along the quantization axis,
are conserved to a good approximation.
It is therefore possible to
label the unperturbed quantum states for the atomic-clock atoms
by the quantum numbers $\ket{F,m_F}$.
The energy-level shift for the state $\ket{F,m_F}$,
calculated in the laboratory frame
with 3-coordinate taken as the quantization axis,
is
\bea
\de E(F,m_F) &=&
\bra{F,m_F} \pr{h} \ket{F,m_F}
\nonumber \\
&=& \hmf \sum_w
(\be_w\tb_3^w + \de_w\td_3^w + \ka_w\tg_d^w)
+ \tmf \sum_w (\ga_w\tc_q^w + \la_w\tg_q^w) \ .
\label{AtomicShift}
\eea
In this expression,
the quantities
$\be_w$, $\de_w$, $\ka_w$, $\ga_w$, and $\la_w$
are expectation values of combinations
of spin and momentum operators
calculated using the extremal states
$\ket{F,m_F=F}$.
The quantities $\hmf$ and $\tmf$ are particular ratios
of Clebsch-Gordan coefficients.
The calculations of quantities appearing in
\rf{AtomicShift} are only approximate,
depending on the nuclear model used.
The quantities with tildes
are specific combinations of Lorentz-violation parameters
that are the only possible parameter combinations
to which clock-comparison experiments are sensitive.
There are five such combinations;
for example, $\tb_3^w$ is given by
\beq
\tb_3^w := b_3^w -m_w d_{30}^w
 + m_w g_{120}^w -H_{12}^w
 \quad.
\eeq
More details of these quantities can be found in Ref.~[\refcite{cc}].
Once the energy-level shifts are known,
the effects of Lorentz violation on the clock frequency
for the transition $(F,m_F) \rightarrow (\pr{F},\pr{m}_F)$
can be found from the difference
\beq \de\om = \de E(F,m_F)- \de E(\pr{F},\pr{m}_F)
\quad .
\label{FrequencyComparison}
\eeq

\section{The Standard Inertial Reference Frame}
The SME framework shows that Lorentz violation could occur in nature
in the form of a variety of background
{\em observer} Lorentz tensors.
Experimental Lorentz tests aim to
measure these tensor components
using a reference frame that is unavoidably noninertial.

Since different experiments use different laboratory coordinates,
it is important to standardize the reference frame
in which the measurements are reported
so that comparisons can be made.
The conventional reference frame that has been used
has origin at the center of the Sun,
and $Z$ axis running northwards parallel to the Earth's rotation axis.
The $X$ axis points towards the vernal equinox on the celestial sphere,
and the orthogonal right-handed system is completed by the appropriate choice of $Y$ axis.
The standard time coordinate, denoted by $T$,
is measured by a clock at the center of the Sun
with time origin at the vernal equinox in the year 2000.
This frame is approximately inertial over periods of thousands of years.
We note that the choice of an Earth-centered frame would be approximately inertial
only for periods on the order of 10 days.

Any given experiment is conducted in a laboratory
with spatial coordinates $(x_1, x_2, x_3)$,
where the third coordinate is defined to be the quantization axis.
This frame may be on the Earth or in space,
and is not generally inertial.
To obtain the experimental result in the standard frame,
details of the laboratory trajectory in spacetime
need to be included in the analysis.

To illustrate the approach for the case of a satellite,
we consider the superposition of two
circular motions, one being that of the Earth around the Sun,
and the other being that of the satellite around the Earth.
The plane of the Earth's circular motion is inclined
at angle $\et \approx 23^{\circ}$ from the equatorial plane,
and its path intersects the positive $X$ axis at the vernal equinox.
To specify the satellite orbital plane,
we use the angle of inclination $\ze$ relative to the $Z$ axis,
and the right ascension $\al$ of the point on its orbit where it intersects
the equatorial plane in the northward direction.
The experimental laboratory can itself be oriented in
various ways within the satellite
so,
for definiteness,
we choose the $x_1$ axis to point towards the center of the Earth,
and the quantization axis $x_3$ to point along the
velocity vector of the satellite relative to the Earth.
Using appropriate rotation and boost matrices,
results in the laboratory frame
can be transformed into the inertial frame.
As an example, the expression $\tb_3$ has the form:

\begin{footnotesize}
\bea \tb_3 &=&
\codt \Big\{ \Big[
 -\tb_X \sin\al\cos\ze + \tb_Y\cos\al\cos\ze + \tb_Z\sin\ze
 \Big]
+\be_\oplus[\mbox{seasonal terms}\ldots] \Big\}
\nonumber \\
&+& \sodt \Big\{ \Big[
 -\tb_X \cos\al - \tb_Y \sin\al
 \Big]
 +\be_\oplus[\mbox{seasonal terms}\ldots] \Big\}
 \nonumber \\
&+& \ctodt \Big\{
 \be_s[\mbox{constant terms}\ldots] \Big\}
+ \stodt \Big\{
 \be_s [\mbox{constant terms}\ldots] \Big\}
 \nonumber \\
&+&
 \be_s[\mbox{constant terms}\ldots ] \ .
\label{Explicitb3}
\eea
\end{footnotesize}

The satellite orbital frequency $\om_s$ appears as expected,
and we note that signals appear also at twice this frequency.
The much lower orbital frequency of the Earth, $\Om_\oplus$,
appears in the seasonal terms.

Since the transformations between the laboratory and inertial reference frames
include a boost, the appearance of the
speed $\be_\oplus \approx 10^{-4}$ of the Earth relative to the Sun
is as might be expected.
So too, the speed of the satellite relative to the Earth is expected;
in  the case of the International Space Station, this speed is $\be_s \approx 10^{-5}$.
For further details of $\tb_3$,
see Ref.~[\refcite{cc:space}].
The energy shifts appearing in Eq.~\rf{AtomicShift}
depend also on the particular atoms of the clock
and the choice of transition used.
Transitions that change the component of the angular momentum
along the quantization axis are often
the most favorable for Lorentz tests.\cite{hbar}

\section{Microwave and Optical Cavities}
Experiments involving optical and microwave cavities
have attracted much recent interest.
Typically, the centimeter dimensions of microwave cavities are
approximately equal to the wavelength of the radiation,
while optical cavities have much higher frequencies and
correspondingly smaller wavelengths.

To perform calculations for these systems,
we need to consider the pure-photon sector
of the SME.
Considering only renormalizable terms,
which are constructed from operators
that have mass dimension four or less,
the lagrangian is\cite{SME:minkowski}
\beq
\cl  =  -\frac 1 4 F_{\mu\nu}F^{\mu\nu}
+\frac 1 2 (\kaf)^\ka\ep_{\ka\la\mu\nu}A^\la F^{\mu\nu}
- \frac 1 4 \kfi F^{\ka\la}F^{\mu\nu} ,
\label{lagrangian}
\eeq
with $F_\mn \equiv \prt_\mu A_\nu -\prt_\nu A_\mu$.
Conventional photon physics is contained in the first term,
while the second and third terms
describe Lorentz-violating interactions.
They contain the coefficients for Lorentz violation
$(\kaf)^\ka$ and $\kfi$,
which are CPT odd and CPT even respectively.
The coefficient $\kfi$ has
the symmetries of the Riemann tensor
and a zero double trace,
giving it a total of 19 independent components.

The second term has been studied extensively,\cite{radiative_corr}
and since it has been exceptionally tightly constrained
by polarization data from distant astronomical sources,\cite{em:theor}
$(\kaf)^\ka$ is set to zero.
A number of the components of the third term
have also been tightly constrained by optical data
from distant cosmological sources.\cite{km1}
The remaining components, nine in total, have been the focus
of a variety of experimental tests with cavities in the last few years.

In cavity oscillators,
the resonant-frequency shift
$\de\nu/\nu$
is the important experimental quantity.
For a cavity
with resonant angular frequency $\om_0$,
we take
$\vec E_0$, $\vec B_0$, $\vec D_0$, and $\vec H_0$
to be the conventional fields.
These can be perturbed by nonzero $\kf$ coefficients,
giving perturbed fields $\vec E$, $\vec B$, $\vec D$, and $\vec H$
with resonant frequency changed by
$\de\nu = \de\om /2\pi \, .$
The Lorentz-violating Maxwell equations
derived from \rf{lagrangian}
lead to the following
expression for the fractional resonant-frequency shift,
\bea
\fr{\de\nu}\nu&=&
-\left(\int_V d^3x \bigl(\vec E_0^*\cdot\vec D
+\vec H_0^*\cdot\vec B\bigr)\right)^{-1}
\nonumber\\
&&
\times\int_V d^3x \left(\vec E_0^*\cdot\vec D-\vec D_0^*\cdot\vec E
-\vec B_0^*\cdot\vec H+\vec H_0^*\cdot\vec B\right.
\nonumber\\
&&\left.
\qquad
\qquad
-i\om_0^{-1}\vec\nabla\cdot
(\vec H_0^*\times\vec E-\vec E_0^*\times\vec H)
\right) ,
\label{dnu}
\eea
with integrals taken over the volume $V$
of the cavity.
Various assumptions regarding
boundary conditions and cavity losses are made,
and since Lorentz violation can be expected
to be small,
only leading-order effects on $\de \nu / \nu$ need to be considered.
For further details see Ref.~[\refcite{km2}].
In the following,
we consider optical cavities first
and then microwave cavities.

Optical tests of Lorentz invariance started with
the classic tests involving the speed of light
done by Michelson and Morley to search for spatial anisotropy.
Later tests by Kennedy and Thorndyke sought dependence
of the speed of light on the speed of the laboratory.
The SME provides a general framework for
both these types of experiments,
and indeed for experiments from all areas of physics.
In the following,
modern versions of these experiments are considered.

An optical cavity
can be analyzed by considering it to be
two parallel reflecting surfaces
with plane waves traveling between them.
With this approach and some simplifying assumptions,
Eq.\ \rf{dnu}
gives the following
expression for the fractional frequency shift
$\de\nu/\nu$
due to Lorentz-violating effects:
\bea
\fr{\de\nu}{\nu}
&=&-\fr1 {2|\vec E_0|^2}
\bigl[\vec E_0^*\cdot(\ka_{DE})_{\rm lab}\cdot\vec E_0/\ep
-(\hat N \times \vec E_0^*)
\cdot(\ka_{HB})_{\rm lab}\cdot(\hat N \times \vec E_0)\bigr] .
\label{dnuopt}
\eea
Here, $\hat N$ is the unit vector along the cavity axis,
$\vec E_0$ specifies the polarization,
and $\ep$ is the transverse relative permittivity.
The coefficients for Lorentz violation
$(\ka_{DE})_{\rm lab}$
and
$(\ka_{HB})_{\rm lab}$
are particular laboratory-frame linear combinations of the $\kf^{\mu\nu\la\ta}$,
as is more fully explained in Ref.~[\refcite{km2}].

Thus, in the presence of Lorentz violation,
the fractional frequency shift
of an optical-cavity oscillator depends
on the cavity orientation
and on the polarization direction of the light.
In any laboratory,
Earth-based or Space-based,
laser light incident on a cavity
will have fractional frequency shift
\bea
\fr {\de\nu}\nu &=&
-\frac14[ 2(\ka_{DE})_{\rm lab}^{33}/\ep
-(\ka_{HB})_{\rm lab}^{11}
-(\ka_{HB})_{\rm lab}^{22} ]
\nonumber \\
&&-\half(\ka_{HB})_{\rm lab}^{12}\sin2\th
\nonumber \\
&&-\frac14 [(\ka_{HB})_{\rm lab}^{11}-(\ka_{HB})_{\rm lab}^{22}]
\cos2\th  ,
\eea
where 1,2, and 3 are the three orthogonal spatial directions.
In general, the laboratory frame is noninertial,
since it is either on the surface of the Earth, or
on a Space craft.
To analyze the results in a standardized form
as per the conventions of other SME works,
the fractional frequency shift must be expressed in
the Sun-centered celestial equatorial basis
through the use of a suitable coordinate transformation.

Microwave-frequency cavity oscillators
are also capable of excellent Lorentz-symmetry tests.
Superconducting cavity-stabilized oscillators
have been considered for use as clocks on the International Space Station.
Niobium superconducting cavities
have achieved $Q$ factors of $10^{11}$ or better,
and frequency stabilities of $3 \times 10^{-16}$
have been attained.

The fractional resonant-frequency shift $\de\nu/\nu$
for a superconducting microwave cavity of any geometry
can be found from Eq.~\rf{dnu}.
For
a cylindrical cavity of circular cross section
that is evacuated
and operated in the fundamental TM$_{010}$ mode,
the fractional frequency shift takes the form:
\bea
\fr{\de\nu}{\nu} \Bigr\vert_{{TM}_{010}}
&=&
-\frac14\hat N^j\hat N^kR^{jJ}R^{kK}
[3(\tilde\ka_{e+})^{JK}+(\tilde\ka_{e-})^{JK}]
\nonumber \\
&&
-\half (\de^{jk}+\hat N^j\hat N^k ) R^{jJ}R^{kK}\ep^{JPQ}\be^Q
\nonumber \\
&&\qquad \times
[3(\tilde\ka_{o-})^{KP}+(\tilde\ka_{o+})^{KP}]
-\tilde\ka_{\rm tr}.
\label{dnuTM010}
\eea
By convention,
lower-case Roman letters denote the laboratory frame
and upper-case ones denote the inertial frame;
$R^{jJ}$ is the spatial rotation from the Sun-centered frame
to the laboratory frame,
$\hat N$ is a unit vector parallel
to the symmetry axis of the cavity,
and $\be^Q$ are the components of the boost vector
in the inertial reference frame.
The various $\tilde \ka$ quantities are coefficients for Lorentz violation
constructed from particular linear combinations of the $\kfi$ quantities.
The above equation applies for experiments in any laboratory.
For a specific experiment,
a variety of parameters are substituted
to include information such as
the geometry of the cavities used,
the trajectory of the laboratory,
and the materials in the system.
The analysis proceeds in a manner similar
to that for clock-comparison experiments,
using transformations to produce the results in the standard reference frame.
One of the advantageous configurations
involves cavities oriented perpendicular to each other.
For more details on these issues, see Ref.~[\refcite{km2}].

\section{The Gravitational sector}
The gravitational sector of the Standard-Model Extension
consists of a framework for addressing Lorentz and CPT violation
in Riemann-Cartan
spacetimes.\cite{gr1}
The usual Riemann and Minkowski
geometries are recovered as limiting cases.
A special case of interest is the
quantum electrodynamics extension in a Riemann-Cartan background.
The coefficients for Lorentz violation
($a_\mu, b_\mu, \ldots$)
typically vary with position,
and the fermion and photon sectors couple to these coefficients.
To obtain the full Standard-Model Extension,
one needs to consider the actions for
the conventional Standard Model,
the Lorentz- and CPT-violating terms built from conventional Standard-Model fields,
and the pure-gravity sector.
Each of these is modified by
gravitational couplings appropriate for a background
Riemann-Cartan spacetime.
This approach allows identification of
the dominant terms in the effective low-energy action for the gravitational sector,
thereby completing the formulation of the leading-order
terms in the SME with gravity.
In principle,
Lorentz symmetry breaking can occur in one of two ways:
it must either be explicit in the action,
or it must occur in the solutions obtained from an action that is itself Lorentz symmetric.
In other words, Lorentz violation must be either explicit or spontaneous.
One of the profound and surprising results obtained in Ref.~[\refcite{gr1}]
is the incompatibility of explicit breaking with generic Riemann-Cartan geometries.
No such difficulties occur in the case of spontaneous Lorentz violation.

The Nambu-Goldstone theorem basically states that a massless particle arises
whenever a continuous global symmetry of the action is not a symmetry of the vacuum.
This means, then,
that spontaneous Lorentz violation may give rise
to various massless particles, or Nambu-Goldstone modes.
In nature, known massless particles include the photon and the graviton,
both of which have two independent polarizations.
It is natural to ask whether the massless modes
following from spontaneous Lorentz violation
are compatible with the photon and the graviton.
This issue is addressed in Ref.~[\refcite{gr2}].
A variety of interesting issues arise.
One is that
spontaneous violation of Lorentz symmetry
is always associated with spontaneous violation of diffeomorphism symmetry
and vice versa.
Another aspect is the
10 possible Nambu-Goldstone modes
associated with the
six generators for Lorentz transformations and
the four generators for diffeomorphisms.
The fate of these modes has been found to depend on the spacetime geometry
and the dynamics of the tensor field triggering the spontaneous Lorentz violation.

Explicit models involving tensor fields that take on vacuum values
spontaneously can be used to study
generic features of the Nambu-Goldstone modes
in the case of Minkowski, Riemann, and Riemann-Cartan spacetimes.
One example is the so-called
Bumblebee model,\cite{gr2}
involving a vector field $B_\mu$.
This idea of using a potential to spontaneously break Lorentz symmetry,
thus enforcing a nonzero vacuum value for $B_\mu$,
was introduced by Kosteleck\'y and Samuel.\cite{slv2}

Remarkably, in Minkowski and Riemann spacetimes,
the bumblebee model generates a photon as a Nambu-Goldstone boson for spontaneous
Lorentz violation.
In principle,
such theories can be experimentally tested,
because there are unconventional Lorentz-violating and Lorentz-preserving
couplings that could be observed in sensitive experiments.
It has also been shown in Ref.~[\refcite{gr3}]
that in Riemann-Cartan spacetime, the Nambu-Goldstone modes could be
absorbed into the torsion component of the gravitational field
through a Higgs mechanism for the spin connection.

Another study of
the Nambu-Goldstone modes arising from spontaneous Lorentz violation
has been done for the case of a
two-index symmetric `cardinal' field $C_{\mu\nu}$
taking on a vacuum value.\cite{gr3}
As in the case of the bumblebee field,
the result found is intriguing:
two massless modes are generated
that correspond to the two polarizations of the graviton.
At low energy and temperature,
conventional gravity is recovered,
and small experimentally testable differences from conventional gravity.
These testable findings,
that the photon and the graviton could be
evidence of spontaneous local Lorentz and diffeomorphism violation,
hold the potential for a profound impact on conventional physics.

An extensive study of the
pure-gravity sector of the minimal Standard-Model Extension
in the limit of Riemann spacetime
has been performed.\cite{gr4}
Under simple assumptions,
there are 20 Lorentz-violating coefficient fields
that take on vacuum values
and lead to
modified Einstein field equations in the limit of small
fluctuations about the Minkowski vacuum.
The equations can be solved
to obtain the post-newtonian
metric.
This work\cite{gr4} includes
a detailed theoretical investigation
of experimental tests and some estimated bounds
for experiments involving
lunar and satellite laser ranging,
laboratory experiments with gravimeters and torsion pendula,
measurements of the spin precession of orbiting gyroscopes,
timing studies of signals from binary pulsars,
and the classic tests
involving the perihelion precession
and the time delay of light.
The sensitivities in these experiments
range from parts in $10^4$ to parts in $10^{15}$.

\end{document}